\documentclass[10pt,superscriptaddress,nofootinbib,showpacs]{revtex4}
\usepackage{epsf}
\usepackage{epsfig}
\usepackage{anysize}
\usepackage{graphicx}
\usepackage{amssymb,amsmath}
\usepackage{float}
\usepackage{color}
\usepackage{ulem}

\restylefloat{table}

\begin{document}
\title{Inherently trap-free convex landscapes for full quantum optimal control}

\author{Qiuyang Sun}
\affiliation{Department of Chemistry, Princeton University, Princeton, New Jersey 08544, USA}
\author{Re-Bing Wu}
\affiliation{Department of Automation, Tsinghua University and Center for Quantum Information Science and Technology, TNList, Beijing, 100084, China}
\author{Tak-San Ho}
\author{Herschel Rabitz}
\affiliation{Department of Chemistry, Princeton University, Princeton, New Jersey 08544, USA}
\date{\today}

\begin{abstract}
We present a comprehensive analysis of the landscape for full quantum-quantum control associated with the expectation value of an arbitrary observable of one quantum system controlled by another quantum system. It is shown that such full quantum-quantum control landscapes are convex, and hence devoid of local suboptima and saddle points that may exist in landscapes for quantum systems controlled by time-dependent classical fields. There is no controllability requirement for the full quantum-quantum landscape to be trap-free, although the forms of Hamiltonians, the flexibility in choosing initial state { of the controller}, as well as the control duration, can infulence the reachable optimal value on the landscape. All level sets of the full quantum-quantum landscape are connected convex sets. Finally, we show that the optimal solution of the full quantum-quantum control landscape can be readily determined numerically, which is demonstrated using the Jaynes-Cummings model depicting a two-level atom interacting with a quantized radiation field.
\end{abstract}

\maketitle

\section{Introduction}
A large body of quantum optimal control simulations and experiments \cite{rev,cat} involve tailored electromagnetic fields acting on atoms or molecules to control a specified physical and/or chemical process, frequently accompanied by optimization of the performance of the control field \cite{teach,Gross}. Aiming to explain the widely observed experimental successes along with much larger numbers of almost perfect optimal control simulations, rigorous analyses were made in the past decade on the {\it quantum control landscape}, herein defined as the expectation value of a desired system observable as a function(al) of the control \cite{land}. The landscape topology is important for establishing the feasibility of finding globally optimal controls, especially when myopic algorithms (e.g., gradient ascent/descent) are employed. A central concern is the properties of control landscape {\it critical points}, where the landscape gradient with respect to the control is zero. Of particular importance is the appearance of suboptimal local extrema as {\it traps}, which could halt a gradient search and prevent reaching the global optimum.

A general control problem involves a physical system and a control that steers the system's dynamics. Both the system and the control can be either classical or quantum mechanical in nature, giving a tetrad of control scenarios: C-C (a classical system steered by a classical control), Q-C (a quantum system steered by a classical control), C-Q (a classical system steered by a quantum control), and Q-Q (a quantum system steered by another quantum control). Most control studies, as well as the associated landscape analyses, are expressed in the {\it semiclassical} framework, i.e., the Q-C scenario. It has been proved that with unconstrained classical control fields, the landscape for a controllable finite-level quantum system is expected to be free of any local suboptima, with all critical points being the global maximum or minimum, or possibly {\it saddle} points {\cite{land,Ben}}. Inclusion of classical field constraints, for example, on the pulse amplitude, bandwidth, and length, may result in additional topological features on the landscape, including possibly traps \cite{cons}. The origins of the rare traps without field constraints and the frequent presence of saddles have remained a mystery.

In this paper, we consider the Q-Q control landscape by taking a full quantum perspective. Suppose that a quantum system $A$ (herein the system) is controlled by another quantum system $B$ (the {\it control}), which can be arbitrarily manipulated by suitable means \cite{Sun-06,Sun-07,Sun-09,Ales-1,Ales-2,Ales-3}. In principle, both quantum systems may take many forms including quantized radiation fields in an optical cavity \cite{Loudon, Gruebele, Briggs, Deffner} as demonstrated by few-photon pulses in a recent experiment \cite{Giesz}, bulk material systems, spins, etc., forming a diverse picture of Q-Q control scenarios. In this work the control is realized by the initial density matrix of the quantized control $B$ coupled with the system $A$, in contrast to the semiclassical Q-C perspective of time-domain fields serving as the controls directly acting on quantum system $A$. We will demonstrate that the landscape expressed in the Q-Q context is {\it convex}, thus rigorously free of any local traps or saddles without requiring any auxiliary assumptions. Furthermore, we will show that the Q-Q optimal solution can be directly calculated from the eigenvectors of a specified operator matrix encapsulating all relevant physical information. The landscape study in this work subsumes so-called incoherent control \cite{Ales-2}. {This work primarily explores the Q-Q landscape for fundamental insights, including the distinction evident from that of the Q-C landscape. Practical laboratory implementation of Q-Q control is understood to call for a detailed assessment of each particular scenario, which is beyond the scope of the present work. As a remark, at present time Q-C control offers the easiest laboratory scenario but there are already example of Q-Q control pointed out above.}

The remainder of this paper will be organized as follows. In Section \ref{Sec:def}, the Q-Q control landscape is defined and proved to be convex. In Section \ref{Sec:sol}, the properties of the resulting convex optimization problem are analyzed, and Section \ref{Sec:exp} presents an example using the Jaynes-Cummings model as an illustration. Finally, concluding remarks are made in Section \ref{Sec:con}.

\section{The convex full quantum landscape}\label{Sec:def}
In the semiclassical Q-C scenario, the {\it kinematic} control landscape of the observable $O_A$ for a closed $N$-level quantum system $A$ controlled by an external classical field $u(t)$ over the duration $T$ has the {\it cost function}
\begin{equation} \label{UrhoUO}
J_{\rm sc}[U_{\rm sc}(T)]={\rm Tr}[U_{\rm sc}(T)\rho_A(0)U_{\rm sc}^\dag(T)O_A],
\end{equation}
where the propagator $U_{\rm sc}(T)$ {(i.e., with the label `sc' referring to semiclassical)} belongs to the unitary group U($N$) designated as the kinematic control space, and $\rho_A(0)$ is the initial density matrix of $A$. It has been shown that the topology of $J_{\rm sc}[U_{\rm sc}(T)]$ is equivalent to that of the corresponding {\it dynamical} landscape $J_{\rm sc}[u(t)]$, as a functional of the control field $u(t)$, upon satisfaction of three assumptions: (i) the system is {\it controllable} \cite{ctrl}, (ii) the local mapping $\delta u(t)\mapsto\delta U_{\rm sc}(T)$ bridging the kinematic and dynamical landscapes is {\it surjective} at any control \cite{surj,Pechen,Schirmer-1}, and (iii) the control fields $u(t)$ are unconstrained, or in practice have sufficient freedom to exploit assumptions (i) and (ii) \cite{cons}. The landscape topology based on these assumptions have been the focus of many previous studies \cite{land}. We remark that assumptions (i) and (ii) can be shown as being ``almost always'' satisfied \cite{Alta,Ben}, consistent with the semiclassical landscape rarely exhibiting traps, while at most only non-trapping saddles.

In the Q-Q scenario, the total Hamiltonian of the composite system $A/B$ is time-independent and can be generally expressed as
\begin{equation}\label{Htot}
H_{AB}=H_A^0\otimes\mathbb{I}_B + \mathbb{I}_A\otimes H_B^0 + \sum_{k} H_A^k\otimes H_B^k,
\end{equation}
where $\mathbb{I}_A$ ($\mathbb{I}_B$) is the identity operator in the Hilbert space of $A$ ($B$), and $H_A^0$ ($H_B^0$) is the respective uncoupled Hamiltonian of $A$ ($B$). The terms $H_A^k$ and $H_B^k$ are the interaction Hamiltonians associated with $A$ and $B$, respectively. Assuming the composite system $A/B$ to be closed, the total unitary propagator produced by the constant Hamiltonian is simply $U_{AB}(t)=\exp(-\frac{i}{\hbar}H_{AB}t)$. One way of making the semiclassical approximation from this formulation is to treat $H_A^k$ as the control Hamiltonians of the system $A$, and the expectation values of $H_B^k$ in the control $B$ as the corresponding classical fields \cite{Layden}.

The Q-Q control objective, parallel to that of the semiclassical case, aims to optimize the expectation value of an observable $O_A$ at time $T$, initially given the state of $A$ as $\rho_A$. The composite system $A/B$ initially may be prepared (i) in a {\it separable} state $\rho = \rho_A\otimes\rho_B$ at $t=0$, with $\rho_A$ being a counterpart of $\rho_A(0)$ in Eq. \eqref{UrhoUO} and $\rho_B$ the initial state of the control $B$; or (ii) in an entangled initial state $\rho$ with the additional constraint that ${\rm Tr}_B(\rho)=\rho_A$, i.e., the reduced density matrix of $A$ is still $\rho_A$. Once the initial state is created, $A$ and $B$ will freely evolve together under the total Hamiltonian $H_{AB}$ and generally become entangled. The control scheme analyzed here will be feasible if the control $B$ can be prepared at a specified quantum state \cite{arb,arb-2}, prior to its interaction with the system $A$. This scenario can be understood in the context of quantum metrology where the initial quantum state of the probe is selected for improving the precision of measurement {\cite{Lloyd}}.

In particular, for a separable initial state $\rho_A\otimes\rho_B$, the landscape in the Q-Q formulation can be defined as
\begin{eqnarray}\label{land}
J_Q[\rho_B] & = & {\rm Tr}[U_{AB}(T)(\rho_A\otimes\rho_B)U_{AB}^\dag(T)(O_A\otimes\mathbb{I}_B)],
\end{eqnarray}
where the control $\rho_B$ is a positive semidefinite matrix of trace one. Here we note that $J_Q(\rho_B)$ can be further written as
\begin{equation}\label{}
  J_Q(\rho_B)=Tr{Tr_B{U_AB(T)(\rho_B\otimes\rho_A)U_AB(T)^\dag}O_A} =Tr{\rho_A(T) O_A},
\end{equation}
where $ \rho_A(T)=Tr_B{U_AB(T)(\rho_B\otimes\rho_A)U_AB(T)^\dag}$. Optimization of the corresponding landscape can be posed as a semidefinite programming problem:
\begin{eqnarray} \label{opt1}
& {\mbox{\rm max/min }} & J_Q[\rho_B] \nonumber \\
& {\mbox{\rm subject to: }} & {\rm Tr}\rho_B = 1, \rho_B\succeq0.
\end{eqnarray}
From Eqs. \eqref{land} and \eqref{opt1}, we note that (1) the cost function $J_Q$ is {\it linear} with respect to the control variable $\rho_B$, thus both convex and concave, and (2) the admissible set of $\rho_B$ is a closed convex set. The condition (2) can be understood as follows: a linear combination of arbitrary density matrices $\{\rho_B^{(k)}\}$, $\varrho_B=\sum_k\lambda_k\rho_B^{(k)}$ with nonnegative coefficients $\lambda_k$ that sum to 1, must satisfy that ${\rm Tr}\varrho_B=1$ and $\varrho_B\succeq0$, thus still being a physically allowed density matrix. The convexity of $\rho_B$ may be violated by additional restrictions on the available initial states of the control $B$, whose analysis is left for future studies.

With the conditions (1) and (2) above, and from the theory of convex optimization \cite{Boyd}, it can be shown that the landscape $J_Q[\rho_B]$ is trap-free, i.e., a local maximum or minimum must also be a global one, if the admissible $\rho_B$ form a convex set. Specifically, for a local minimum $\rho_B^\ast$ of the function $J_Q[\rho_B]$ and an arbitrary $\rho_B'\neq\rho_B^\ast$, the convex combination $(1-\lambda)\rho_B^\ast+\lambda\rho_B'$ with $\lambda\to 0_+$  is within the neighborhood of $\rho_B^\ast$, thus
\begin{eqnarray} \label{proof}
J_Q[\rho_B^\ast] &\le& J_Q[(1-\lambda)\rho_B^\ast + \lambda\rho_B'] \nonumber \\
&=&(1-\lambda)J_Q[\rho_B^\ast] + \lambda J_Q[\rho_B'],
\end{eqnarray}
which immediately leads to the relation $J_Q[\rho_B^\ast] \le J_Q[\rho_B']$, so $\rho_B^\ast$ must also be a global minimum. Similarly it can be proved that a local maximum of $J_Q[\rho_B]$ is also a global maximum.

By the linearity of the function $J_Q$, we further find that all {\it level sets} of the Q-Q landscape must also be convex sets, and thus must be {\it connected}. A {\it level set} is defined by the set of all controls with an identical cost function value, whose topology (especially the {\it connectivity}) was studied in the semiclassical formulation in Ref. \cite{Dominy}. By the linearity of the function $J_Q$, we further find that all level sets of the Q-Q landscape are connected, as shown below. Given two initial states $\rho_{B,1}$ and $\rho_{B,2}$ of the control $B$ which are on the same level set of the Q-Q landscape, i.e., $J_Q[\rho_{B,1}] = J_Q[\rho_{B,2}] = J_0$, then any convex combination of $\rho_{B,1}$ and $\rho_{B,2}$ will also be on the level set at $J_0$ since $J_Q$ is a linear function of $\rho_B$,
\begin{equation}
J_Q[\lambda\rho_{B,1}+(1-\lambda)\rho_{B,2}] = \lambda J_Q[\rho_{B,1}] + (1-\lambda)J_Q[\rho_{B,2}] =  \lambda J_0 + (1-\lambda)J_0 = J_0, \indent \lambda\in[0,1].
\end{equation}
Therefore, the level set $\{\rho_B|J_Q[\rho_B] = J_0\}$ for any reachable $J_0$ value must be a convex set, and thus connected.

The simple convexity of the control landscape $J_Q$, Eq. \eqref{land}, in the Q-Q formulation is in sharp contrast to the semiclassical counterpart $J_{\rm sc}$, Eq. \eqref{UrhoUO}, which is a highly nonlinear functional of the control field $u(t)$. The control landscape features in these two different formulations are summarized in Table \ref{table:1}. Note that there is no controllability \cite{Ales-2,Ales-3} requirement for the Q-Q landscape to be trap-free, which is unnecessary for convexity of the admissible set of the control $\rho_B$, or the cost function $J_Q$. However, the forms of Hamiltonians $H_A^k$ and $H_B^k$ in Eq. \eqref{Htot}, the flexibility in creating $\rho_B$, as well as the choice of final time $T$ can influence the optimal value of $J_Q$ reachable in the Q-Q control scenario.

\begin{table}[H]
\centering
\begin{tabular}{ c | c  c  }
\hline Formulation & Q-C & Q-Q \\
\hline Nature of control & $u(t)$ & $\rho_B$ \\
       Cost function & nonlinear & linear \\
       Landscape topology & trap-free$^\dag$ & convex, trap-free$^\ast$ \\ \hline
\end{tabular}
\caption{Summary of the control landscapes in the Q-C and Q-C scenarios. \\
$^\dag$ Trap-free upon satisfaction of three key assumptions, and possibly with saddles. $^\ast$ No saddles or other suboptimal critical points present.}
\label{table:1}
\end{table}

As an alternative of the main problem in Eqs. \eqref{land} and \eqref{opt1}, if the initial state $\rho$ of $A/B$ is entangled and cannot be separated as $\rho_A\otimes\rho_B$, the landscape in the Q-Q framework can instead be formulated as
\begin{eqnarray}
& {\mbox{\rm max/min }} & J_Q[\rho] = {\rm Tr}[U_{AB}(T)\rho U_{AB}^\dag(T)(O_A\otimes\mathbb{I}_B)] \nonumber \\
& {\mbox{\rm subject to: }} & {\rm Tr}_B(\rho) = \rho_A, \indent {\rm Tr}\rho=1, \indent \rho\succeq0.
\end{eqnarray}

It can be easily verified that this circumstance also entails a convex optimization problem, being a linear cost function in a convex admissible set of $\rho$, and thus the landscape $J_Q[\rho]$ is also free of local traps.

\section{Optimal solutions for the full quantum landscape}\label{Sec:sol}
A complete optimal solution for the landscape \eqref{opt1} in the Q-Q formulation can be obtained by recasting Eq. \eqref{land} as
\begin{equation}\label{rhoM}
J_Q[\rho_B]={\rm Tr}(\rho_B\mathcal{O}_B)
\end{equation}
where the partial trace over $A$
\begin{equation}\label{M}
\mathcal{O}_B:={\rm Tr}_A[U_{AB}^\dag(T)(O_A\otimes \mathbb{I}_B)U_{AB}(T)(\rho_A\otimes \mathbb{I}_B)]
\end{equation}
gives rise to the {\it landscape observable} associated with the control landscape $J_Q[\rho_B]$ in the Hilbert space spanned by the density matrix $\rho_B$. {All the necessary information, $\rho_A$, $O_A$ and $U_{AB}(T)$, for identifying the landscape optimum resides in the single operator $\mathcal{O}_B$, which plays the role of an effective ``observer'' enabling the landscape with respect to $A$ to be extracted \cite{observer}.} The upper and lower bounds of $J_Q[\rho_B]$ in Eq. \eqref{rhoM} can be given in terms of the eigenvalues of $\mathcal{O}_B$, i.e., $\mathcal{O}_B^{\rm min} \le {\rm Tr}(\rho_B\mathcal{O}_B) \le \mathcal{O}_B^{\rm max}$, where $\mathcal{O}_B^{\rm max}$ and $\mathcal{O}_B^{\rm min}$ are the maximal and minimal eigenvalues of $\mathcal{O}_B$, respectively. To reach the global maximum (minimum) of $J_Q$, $\rho_B$ must be composed of the eigenstate(s) of $\mathcal{O}_B$ corresponding to its maximal (minimal) eigenvalue. For a general degenerate eigenvalue $\mathcal{O}_B^\ast$ ($\ast$ stands for max or min), the landscape optimal solutions
\begin{equation} \label{rhoCopt}
\rho^\ast_B=\sum_ip_i|i\rangle\langle i|, \indent p_i\ge0, \indent \sum_i p_i=1,
\end{equation}
form a convex set of mixed states, with $\{|i\rangle\}$ being the subspace of degenerate eigenstates of $\mathcal{O}_B$ associated with $\mathcal{O}_B^\ast$, thus leading to the optimal cost function value $J_Q[\rho_B^\ast]=\mathcal{O}_B^\ast$. If $\mathcal{O}_B^\ast$ is nondegenerate, the optimal control $\rho_B^\ast=|i\rangle\langle i|$ can only be a pure state, i.e., an extremal point on the boundary of the admissible set. We remark that in general there are infinitely many distinct control fields at the semiclassical dynamical landscape optimum, which require identification by deterministic or stochastic searching algorithms \cite{Schirmer}. In contrast, the optimal solution $\rho_B^\ast$ of the Q-Q control landscape $J_Q$, Eq. \eqref{land}, can be explicitly determined directly from Eq. \eqref{rhoCopt}.

\section{Illustration: full quantum control in the Jaynes-Cummings model}\label{Sec:exp}
We consider the Jaynes-Cummings (JC) model \cite{JCM}, which describes a two-level atom (the target $A$) with a ground state $|g\rangle$ and an excited state $|e\rangle$, interacting with a quantized radiation field (the control $B$) containing a single bosonic mode with countably infinite number states $|n\rangle$, $n=0,1,\cdots$. In the rotating wave approximation, the total Hamiltonian is written as
\begin{equation} \label{JC}
H_{AB}= \frac{\omega}{2}\sigma_z + \nu a^\dag a + \frac{\Omega}{2}(\sigma_+a + \sigma_-a^\dag)
\end{equation}
where $\omega$ and $\nu$ are the frequencies of the atom and the field, respectively, and $\Omega$ is the coupling strength. $a^\dag$ and $a$ are the creation and annihilation operators of the field, while $\sigma_+=|e\rangle\langle g|$, $\sigma_-=|g\rangle\langle e|$, and $\sigma_z=|e\rangle\langle e| - |g\rangle\langle g|$ are operators of the atom. A semiclassical Q-C counterpart of Eq. \eqref{JC} can be formulated as
\begin{equation}\label{sc-2}
H_{\rm sc}(t)=\frac{\omega}{2}\sigma_z + \frac{1}{2} [u(t)\sigma_+ + u^\ast(t)\sigma_-]
\end{equation}
where $u(t)$ is a complex-valued classical field. The landscape maximum $J_{\rm sc}^{\rm max}=1$ can be achieved as well with unconstrained $u(t)$. No saddles are present in this two-level single particle case for system $A$, but multi-level semiclassical cases will generally have landscape saddles; in contrast the Q-Q landscape as strictly convex is always free of hindering saddles.

\begin{figure}
\centering
\includegraphics[trim=0 0 250 100, clip, width=0.48\columnwidth]{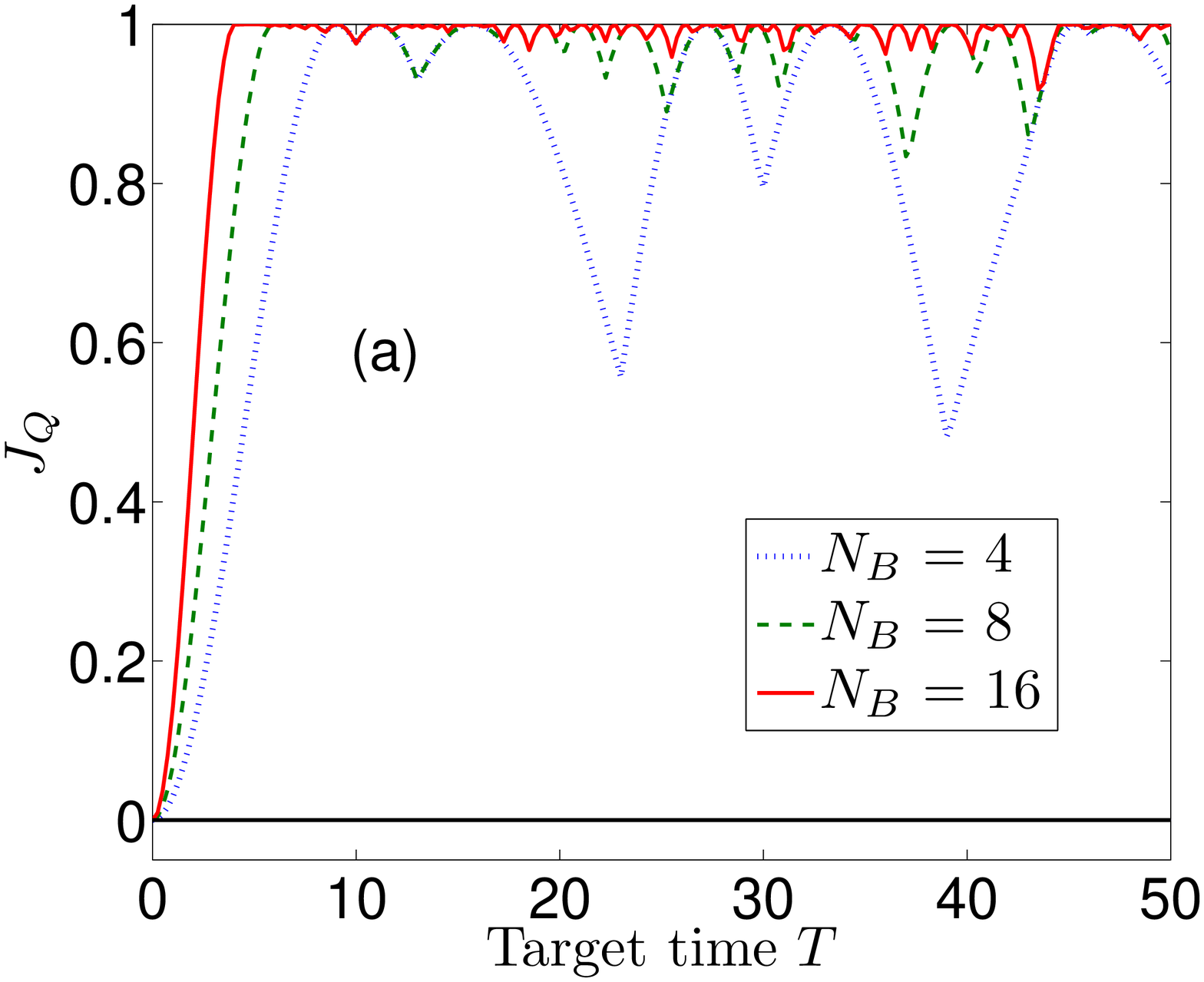}
\includegraphics[trim=0 0 250 100, clip, width=0.48\columnwidth]{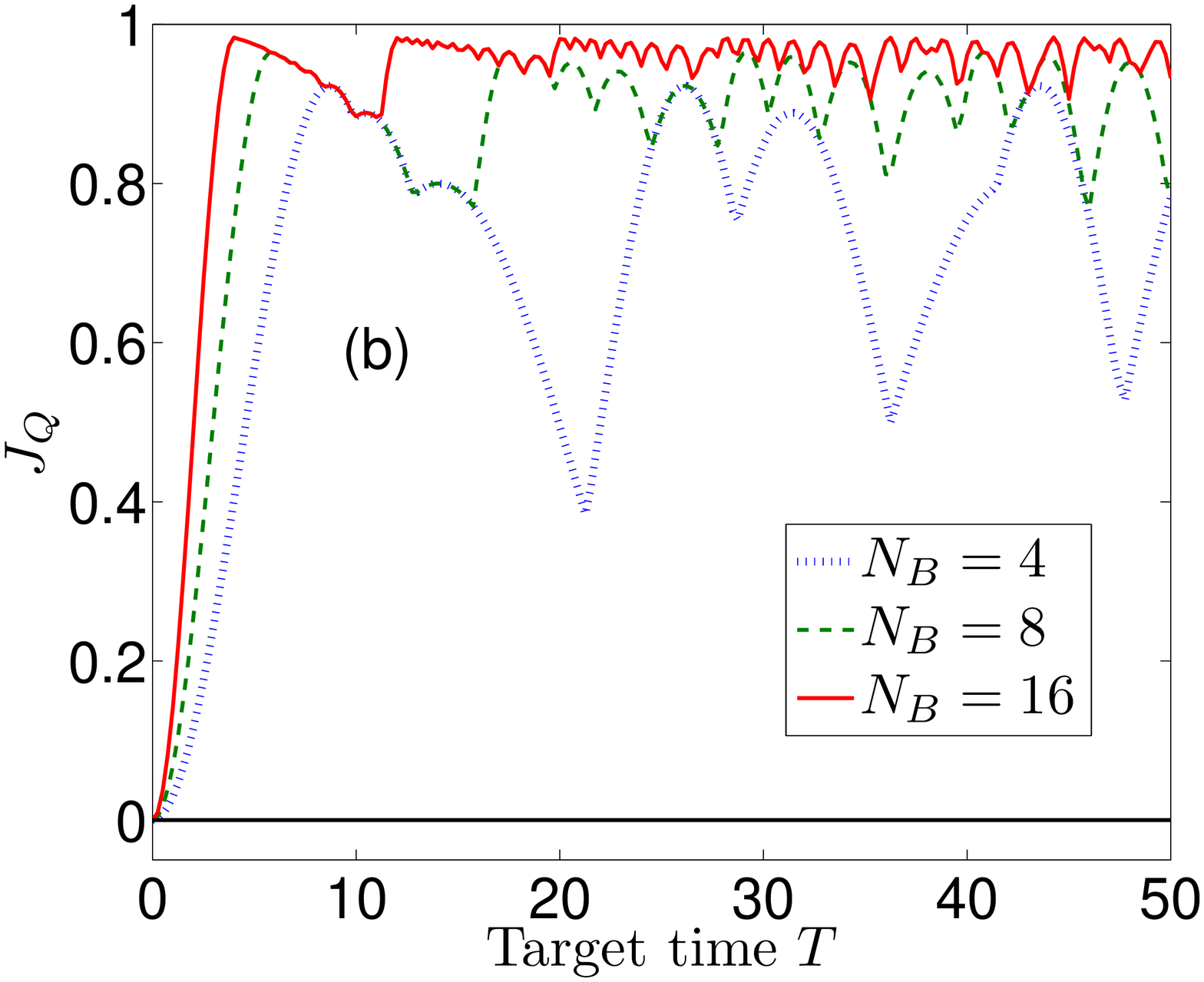}
\caption{(color online). State transition $|g\rangle\to|e\rangle$ in a truncated JC model: upper bound curves of a Q-Q control landscape $J_Q$ at different $T$, with first $N_B$ levels of the quantized field selected as the control resource. The frequency detuning $\Delta=\nu-\omega$ is set to (a) $\Delta=0$ (on-resonance) and (b) $\Delta=0.1$ (off-resonance). Regardless of the value of $J_Q$ the landscape is trap free, and the optimal value of the control $\rho_B^\ast$ may be readily identified, as explained in the text. The lower bound for $J_Q$ is always zero, shown as the flat line. } \label{fig:2}
\end{figure}

Here we consider a control landscape with the form of Eq. \eqref{opt1}, with the initial state of the quantized field as the control $\rho_B$ utilized to optimize the transition probability from the ground to the excited state in the atom, i.e., we specify that $\rho_A=|g\rangle\langle g|$ and $O_A=|e\rangle\langle e|$. Using Eq. \eqref{M}, the resultant $\mathcal{O}_B$ is a diagonal matrix with $\langle 0|\mathcal{O}_B|0\rangle = 0$ and
\begin{equation} \label{eigM}
\langle n|\mathcal{O}_B|n\rangle = \sin^2\alpha_n \sin^2[\frac{T}{2}\sqrt{\Delta^2 + \Omega^2n}]
\end{equation}
for $n=1,2,\cdots$, where $\Delta=\nu-\omega$ is the detuning and $\alpha_n := -\tan^{-1}(\frac{\Omega\sqrt{n}}{\Delta})$. The eigenvalues of $\mathcal{O}_B$ are distributed within the interval $[0,1]$, the maximum and minimum among which will determine the range of the landscape $J_Q$. In the on-resonance case of $\Delta=0$ and thus $\sin^2\alpha_n=1$, for any $T>0$ there exists some $n$ such that $\sin^2[\frac{T}{2}\sqrt{\Delta^2 + \Omega^2n}]$ approaches 1, and the matrix $\mathcal{O}_B$ has an eigenvalue of 1, which means that full transition from $|g\rangle$ to $|e\rangle$ can be accomplished by the control $\rho_B^\ast=|n\rangle\langle n|$. In the off-resonance case that $\Delta\neq0$, however, we observe that the upper bound for the eigenvalues of $\mathcal{O}_B$ is always less than 1 for any finite $n$, since $|\alpha_n|<\pi/2$. Therefore, the full transition may only be asymptotically approached in the limit that $n\to\infty$, i.e., at infinite field strength. In numerical simulations we truncated the first $N_B$ levels of the quantized field, $|n\rangle$ with $n=0,1,\cdots,N_B-1$, and calculated the bounds of the landscape $J_Q$ at different $T$ (see Fig. \ref{fig:2}). The parameters are set to $\Omega=0.2$ and $\Delta=0$ or 0.1 to represent the on- or off-resonance cases. The restriction on the control space makes the full state transition unreachable for most values of the time $T$, but the impact becomes less significant as $N_B$ increases from 4, 8 to 16. A resonant quantized field exhibits better performance than a detuned one at the same level of truncation. The sharp changes of $J_Q$ evident in Fig. \ref{fig:2} arise from $\rho_B^\ast$ jumping from one solution $|n\rangle\langle n|$ to another $|n'\rangle\langle n'|$ as $T$ varies.

\section{Conclusions}\label{Sec:con}
In conclusion, this paper provides a full Q-Q formulation for the control landscape aiming to optimize the expectation value of an observable associated with the system. The system $A$ and the control $B$ are both treated quantum mechanically, which together undergo free evolution governed by a constant total Hamiltonian of the coupled bipartite system. The control consists of the initial density matrix $\rho_B$ of $B$, which may be prepared by any available means. Within this framework, optimization over the landscape $J_Q[\rho_B]$ with respect to the density matrix $\rho_B$ presents a convex problem with a convex admissible set of controls. Therefore, the full Q-Q control landscape is rigorously free of any suboptimal local extrema as either traps or saddles, if no additional constraints are imposed on $\rho_B$ to violate its convexity. The mathematical simplicity of the full Q-Q control problem explicitly permits {readily finding} the landscape optimal solutions $\rho_B^\ast$, and we show that the landscape optimum can always be achieved by some pure (or mixed, as appropriate) initial state $\rho_B^\ast$ of the control. The conclusions here imply that the search for optimal solutions over a full Q-Q control landscape in the laboratory will be efficient provided that an appropriate initial state of the control can be prepared.

In the most general tetrad of control pictures, this work explores the Q-Q picture {as a complimentary} in addition to
existing studies in the Q-C picture. The characteristic control landscapes in the remaining two frameworks (i.e., C-C and C-Q) are
rooted in the fundamental differences of quantum and classical mechanics. These additional landscapes are of theoretical and
practical importance, with the C-C picture only partially explored to date \cite{Joe}, while the
C-Q picture has yet to be physically defined. Since classical dynamics can be taken as a limiting
process of quantum dynamics, the present Q-Q landscape analysis may provide a foundation for future
research to draw together the full tetrad of classical and quantum mechanical control in a seamless fashion \cite{Briggs}.

\begin{acknowledgments}
H.R. acknowledges support from NSF Grant No. CHE-1058644, T.S.H. for the DOE Grant No. DE-FG02-02ER15344, and Q.S. for the ARO Grant No. W911NF-13-1-0237 as well as the Princeton Plasma Science and Technology Program. R.B.W. acknowledges support from NSFC Grants No. 61374091 and No. 61134008.
\end{acknowledgments}

\end{document}